# The Configuration of the Perivascular System Transporting Macromolecules in the CNS

(PREPRINT)

**Beata Durcanova[1], Janine Appleton[1], Nyshidha Gurijala[1], Vasily Belov[1-3], Pilar Giffenig[1], Elisabeth Moeller[1], Matthew Hogan[1], Fredella Lee[1], and Mikhail Papisov[1-3]**

[1] Massachusetts General Hospital, Boston, MA

[2] Harvard Medical School, Boston, MA

[3] Shriners Hospitals for Children – Boston, Boston, MA

**\* Correspondence:**

Mikhail Papisov
papisov@helix.mgh.harvard.edu

**Abstract**

Large blood vessels entering the CNS are surrounded by perivascular spaces that communicate with the cerebrospinal fluid and, at their termini, with the interstitial space. Solutes and particles can translocate along these perivascular conduits, reportedly in both directions. Recently, this prompted a renewed interest in the intrathecal therapy delivery route for CNS-targeted therapeutics. However, the extent of the CNS coverage by the perivascular system is unknown, making the outcome of drug administration to the CSF uncertain.

We traced the translocation of model macromolecules from the CSF into the CNS of rats and non-human primates. Conduits transporting macromolecules were found to extend throughout the parenchyma from both external and internal (fissures) CNS boundaries, excluding ventricles, in large numbers, on average ca. 40 channels per mm$^2$ in rats and non-human primates. The high density and depth of extension of the perivascular channels suggest that the perivascular route can be suitable for delivery of therapeutics to parenchymal targets throughout the CNS.

# 1 INTRODUCTION

The major problem in developing therapies for diseases involving the CNS is the lack of safe and efficient means to overcome the blood-brain, blood-cerebrospinal fluid and blood-arachnoid barriers.[1] It has been estimated that for greater than 98% of small molecules and for nearly 100% of large molecules the brain is not accessible systemically.[2] The exclusion of macromolecules from the CNS is particularly trying because biopharmaceuticals, all of which are large molecules or supramolecular constructs (proteins[3,4,5], oligonucleotides[6,7], gene vectors[8,9,10]), could be particularly potent as therapies for many diseases involving the CNS[11,12,13].

Several avenues of drug delivery to the CNS have been investigated over the last two decades[14], including direct transcranial infusion[15] and transport across the blood-brain barrier using endogenous receptors transporting their ligands (insulin[16], transferrin[17], lipoproteins[18]) across the brain endothelium. Although some preclinical results appear promising[19], these studies have not yet resulted in clinically feasible solutions.

Accumulating evidence suggests that highly potent biopharmaceuticals can exert measurable biological effects in the CNS after administration to the cerebrospinal fluid (CSF)[20,21,22,23,24,25,26], which renewed the interest in the intrathecal (IT) drug delivery route. Presently, the IT route is primarily employed in pain and spasticity management, where bioactive molecules are delivered to the vicinity of nerve roots and the spinal cord, from which they can reach their respective receptors by diffusion.[27] Biopharmaceuticals cannot freely diffuse from the CNS surface deep into the CNS parenchyma; their diffusion from the CSF would unlikely result in significant penetration beyond a few tenths of a millimeter[28,29]. To date, the only plausible explanation of macromolecule entry into the CNS appears to be through the perivascular spaces.

The CSF has long been known to communicate with perivascular spaces.[30] Peroxidase administered to the CSF was found not only in the spaces surrounding large blood vessels (Virchow-Robin spaces) but in the smaller spaces along their branches as well[31]. The available data[32,33,34,35,43] suggest that perivascular solute entrance into the CNS is by 2-3 orders of magnitude too fast to be diffusion-driven and is likely facilitated by pulsation of the blood vessels contained in the space[36,37]. Pulsation can remix the perivascular space longitudinally or cause solute spread via Taylor dispersion[38,39] (similarly, pulsatile waves transmitted from large vessels[40] can facilitate solute spread in the CSF[33]). An alternative mechanistic hypothesis[32] is being debated[41]. The distal termini of perivascular spaces lack continuous walls and communicate with the interstitial fluid[42], which can explain both the observed macromolecule accumulation in a variety of parenchymal cells[43] and the reported biological effects[20-26].

If macromolecule transport from the CSF into CNS is indeed predominantly perivascular, the ubiquity of perivascular channels capable of transporting macromolecules is critical for the accessibility of parenchymal targets. Therefore, we set out to determine the distribution and configuration of the perivascular pathways transporting macromolecules.

Although the walls of the perivascular channels can potentially be stained with a variety of dyes, small molecules administered to the CSF are rapidly cleared to the systemic circulation[44] and only a minute fraction of them can enter the perivascular space (e.g., for left-rotating [$^{18}$F]fluorodeoxyglucose the half-clearance time in monkeys is 15 minutes[45]). A variety of fluorescent macromolecular probes (e.g., lectins) can stain the walls of the perivascular spaces, but the brightness of the tracer layer is not sufficient for reliable channel counting.

We identified a protein avidly taken up by actively endocytosing cells (perivascular macrophages[31,46]) residing in the perivascular spaces. The protein, highly glycosylated for enhanced endocytosis alpha-N-acetyl glucosaminidase (rhNAGLU)[47] conjugated with a fluorophore, was administered into the cisternal CSF of rats and rhesus monkeys. After the uptake, rhNAGLU accumulated in large vesicular compartments of the perivascular macrophages, thus labeling the macromolecule-transporting conduits. The patterns of the tracer deposition in the CNS were investigated by fluorescence microscopy in unfixed unstained cryosections.

## 2 MATERIALS AND METHODS

### 2.1 Preparation of fluorescent tracers.

Proteins were labeled with activated fluorescent dyes (Alexa series, Texas Red and Fluorescein isothiocyanate) from Thermo/Molecular probes. The labeling procedures were carried out generally in accordance with the manufacturer's recommendations, with optimization of protein:dye ratio for maximal fluorescence intensity per protein mass. Fluorescence was measured on a photon counting spectrometer (PTI Quantamaster, Photon Technology International/Horiba Scientific, Edison, NJ) at pH=7.4 for a 10 µl aliquot of the test solution in 3 ml of 0.15 M NaCl buffered with 50 mM sodium phosphate buffer solution, pH = 7.4).

N-acetylglucosaminidase alpha (rhNAGLU) was kindly provided by Synageva Biopharma (presently Alexion, New Haven, CT). Activated fluorescent dyes were purchased from Molecular Probes/ThermoFisher Scientific, Waltham, MA. Other reagents and salts, analytical ACS grade or higher, were from ThermoFisher Scientific and from Sigma-Aldrich, St. Louis, MO.

For the main experimental and control groups, rhNAGLU was labeled with fluorescein isothiocyanate (FITC). One ml of (9.2 mg/ml protein solution was mixed with 0.5 ml of 1 M sodium carbonate buffer, pH=9.4. Under stirring, 40 µl of 67 mg/ml FITC solution in dry DMSO were added by 5 µl aliquots at ambient temperature. After an overnight incubation under stirring at ambient temperature, the protein was isolated by size exclusion chromatography on Sephadex G25 equilibrated with 0.15 M sodium chloride buffered with 20 mM sodium phosphate, pH=7.4. The entire high molecular weight fraction was collected (total volume 1.6 ml, protein concentration: 5.75 mg/ml). For a series of four analogous syntheses, dye content, as determined from the optical densities at 280 and 495 nm, was 5±1 moiety per protein molecule. Analogous synthesis using Alexa 350 NHS ester at pH=9.0 resulted in a product labeled with 8±2 dye moieties per molecule. Using calibration curves of dye emission vs. concentration (excitation 490 and 335 nm, emission 520 and 445 nm, respectively), it was found that Alexa 350 was practically unquenched (90±5% vs. the free hydrolyzed dye fluorescence), whereas the emission of the protein adduct with multiple FITC residues was significantly reduced (likely self quenched), 20±2% as compared to a free fluorescein standard. As a result, the labeled products had similar brightness (photon•sec$^{-1}$•protein molecule$^{-1}$), which is in agreement with the ca. five-fold lower extinction coefficient of Alexa 350 at the excitation maximum, as compared to FITC. For animal studies, the solutions were filtered through a 0.22 μm sterile membrane.

## 2.2 Preparation of radiotracers.

For labeling with Iodine-124, the solutions of macromolecules, 10 to 25 mg/ml in 0.25 M sodium phosphate buffer (0.1 ml), were incubated in Pierce Iodogen iodinating tubes (ThermoFisher Scientific, Waltham, MA) with 0.1-3 mCi (0.01-0.3 ml) of [$^{124}$I]NaI solution containing $10^{-3}$M NaOH (as received from IBA Molecular (presently Zevacor, Dulles, VA) for 40 min. The product was desalted on Sephadex G-25 (PD-10 column, GE Healthcare Life Sciences, Marlborough, MA) equilibrated with sterile saline and analyzed by HPLC (BioRad BioSil 125 column) with UV and gamma detection. Radiochemical yield was 60-80%, depending on the macromolecule. The radioactivity was measured on an Atomlab 100 dose calibrator (Biodex Medical Systems, Shirley, NY).

For labeling with Zirconium-89, macromolecules were first conjugated with 1-(4-isothiocyanatophenyl) deferoxamine (Macrocyclics, Plano, TX) in 0.1 M sodium carbonate buffer, pH=9, at 10 and 1 mg/ml, respectively. After an overnight incubation at ambient temperature, the reaction mixture was spun at 9,000 g for 10 minutes, and the supernatant was desalted and analyzed as described above. [$^{89}$Zr]Zr oxalate in 1M oxalic acid from IBA Molecular (presently Zevacor, Dulles, VA) was neutralized with 2 M $Na_2CO_3$ and then reacted with the deferoxamine-macromolecule conjugates in 0.2 M HEPES buffer, pH=7.1 at ambient temperature for 20 minutes. The products were desalted on PD-10 columns equilibrated with 0.9% NaCl (radiochemical yield: 70-90%).

## 2.3 Animal studies.

All animal studies were carried out in accordance with protocols approved by the Massachusetts General Hospital and in accordance with relevant guidelines and regulations.

Protein circulation was studied by positron emission tomography (PET) after intravenous (IV) or lumbar intrathecal (IT) administration of $^{89}$Zr or $^{124}$I labeled rhNAGLU. Perivascular delineation was studied using fluorophore-labeled rhNAGLU. All procedures were carried out under Isoflurane/0$_2$ anesthesia, in monkeys with continuous monitoring of heart and breathing rates.

Tracer circulation. Tracer circulation studies were carried out by PET to investigate tracer transport in vivo and to estimate tracer catabolization rates after uptake by cells in normal Sprague Dawley CD rats (n=4 for $^{124}$I labeled rhNAGLU, N=6 for $^{89}$Zr labeled rhNAGLU and in non-human primates (*M. fascicularis*), N=4 for $^{89}$Zr labeled rhNAGLU and N=1 for $^{124}$I labeled rhNAGLU.

IV administration of radiolabeled rhNAGLU in both rats and monkeys was carried out through a temporary intravenous catheter installed in the tail or saphenous vein, respectively.

IT administration of radiolabeled rhNAGLU for circulation studies in *M. fascicularis* was carried out through a pre-installed subcutaneous port equipped with a catheter opening into the CSF at L1-L2.

Radiotracer administration was carried out after the initiation of dynamic PET/CT data acquisition, which continued for 30 minutes after the injection. Then, whole body static PET/CT images were acquired at multiple time points, 1 to 72 hours.

The administered doses of rhNAGLU varied from 1 to 20 mg/kg IV in both rats and monkeys, and 3 mg/animal IT in monkeys. The radionuclide doses were 50 to 200 µCi per animal in rats and 0.4 to 1 mCi per animal in monkeys.

Imaging was carried out using a custom PET/CT imaging system consisting of MicroPET Focus 220 PET scanner (Siemens Medical Solutions USA, Inc., Malvern, PA) and CereTom NL 3000 CT scanner (Neurologica, MA). The imagers were aligned and equipped with a custom imaging bed extending through both imagers along the alignment axis, ensuring reliable PET/CT image registration[48]. Focus 220 works in 3D mode and features a 22 cm animal opening, axial field of view (FOV) 7.6 cm and transaxial FOV 19 cm. The scanner's detection system provided a 2.1 mm spatial resolution for $^{124}$I and 1.7 mm spatial resolution for $^{89}$Zr. The energy window of the PET imager was set for the entire study to 350-650 keV, and the coincidence timing window was set to 6 ns. CereTom NL 3000 is a 6-slice tomograph with high-contrast resolution of 0.4 mm (developed for human head imaging in ICU). The image acquisition settings were: tube voltage 100 kV, tube current 5 mA, resolution 6 s/projection, axial mode with slice thickness of 1.25 mm. Image pixel size was set to 0.49x0.49x1.25 mm. The image sharpness was optimized to soft tissue. CT images were used both for anatomical reference and attenuation correction of the PET images. PET/CT image co-registration was carried out manually using ASIProVM software (Siemens/CTI Concorde Microsystems, Knoxville, TN).

PET raw data acquisition and histogramming were carried out on a Dell Precision PWS690 Workstation (Dell, Inc., Round Lake, TX; 3 GB RAM and 4 Xeon 3.20 GHz processors running under a 32-bit Windows XP [Microsoft Corp., Redmond, WA]) using Siemens MicroPET Firmware/Software, release 2.5 (Siemens Medical Solutions, Inc., Malvern, PA). Image reconstruction was carried out on a raid 5 server (17.9 GB RAM and 8 Xeon 2.4 GHz processors running under Microsoft Windows Server 2003, Enterprise x64 Edition) using Siemens MicroPET Firmware/Software, release 2.4.5. All subsequent image processing and analysis were performed on non-host workstations using the ASIProVM software (Siemens/CTI Concorde Microsystems, Knoxville, TN) running under 32-bit Windows XP and Inveon Research Workplace 3.0 (Siemens Medical Solutions, Inc., Malvern, PA) running under 64-bit Windows XP. During raw data histogramming and image reconstruction, the corrections for isotope decay, detectors dead-time, random coincidences, and tissue attenuation were applied. The data were reconstructed into the image matrix with the pixel size of 0.95 mm and fixed slice thickness of 0.8 mm using a 3-dimensional (3-D) ordered-subset expectation maximization/maximum *a posteriori* (OSEM3D/MAP) protocol with the smoothing resolution of 1.5 mm, 9 OSEM3D subsets, 2 OSEM3D and 15 MAP iterations. The data were also reconstructed with Fourier rebinning 2-D filtered backprojection (FORE-2DFBP) [49] to ensure that the numerical data derived from OSEM3D/MAP and FORE-2DFBP reconstructed images were identical and thus excluded possible reconstruction artifacts (none were identified). FORE-2DFBP was performed with a ramp filter cutoff at the Nyquist spatial sampling frequency (0.5 mm$^{-1}$). Whole body images were composed of the acquired single bed position images with a 12 mm overlap.

The dynamic and static PET images were analyzed to obtain numerical kinetic data for multiple manually selected regions of interest (ROIs), including CSF reservoirs, whole brain, grey and white matter areas, cardiac blood pool and major organs. Catabolization (lysosomal depolymerization) rates of rhNAGLU were estimated by clearance of $^{124}$I from [$^{124}$I]I- rhNAGLU accumulating tissues (which is preceded by lysosomal protein hydrolysis followed by microsomal deiodination of iodotyrosine[50]).

Perivascular delineation studies with fluorophore-labeled rhNAGLU were carried out in rats and non-human primates.

Prior to the study, rats from different vendors and of different ages were compared for CNS autofluorescence. The differences were found to be insignificant, with a somewhat higher incidence of focal deposition of material fluorescent in both green and red channels (consistent with lipofuscin) in older animals. In the main experimental and control groups, animals were from Charles River Laboratories (Shrewsbury, MA, USA), Sprague Dawley CD rats, 330±80 g (N=6 per group in the main experimental and control groups).

FITC labeled rhNAGLU was administered IT via direct injection into the cisterna magna. Prior to the injection, rats were inducted with 5.0% of 300 ml/min isoflurane, (Forane, Baxter Healthcare Corporation, USP). During the injection, the rats were given isoflurane at 1.9-2.5% through a facemask. FITC labeled rhNAGLU, 50 µl of a 5.75 mg/ml tracer solution in sterile saline, was injected intrathecally through a 23G needle equipped with a catheter into the cisterna magna. The catheter line (total internal volume <10 µl) was then flushed with 20 µl of sterile saline. Rats with signs of damage within the intrathecal space (blood in the CSF, sharply asymmetrical tracer accumulation in the pia mater) and/or intraparenchymal injection (fluorescence in the longitudinal axonal tufts in the brainstem) were excluded from the studies.

Two female *M. Mulatta* scheduled for euthanasia were used to verify the presence of the same perivascular structures as found in rodents. Seventeen years old primates were selected to determine whether the perivascular cells remain highly endocytosing with age.

Alexa 350 was used in the monkey studies as a fluorophore instead of fluorescein based on the pilot data showing widespread lipofuscin autofluorescence in older primate brain; the emission of Alexa 350 ($Em_{max}$=445 nm) was found to be outside of the spectrum of lipofuscin autofluorescence. IT administration of Alexa 350 labeled rhNAGLU was carried out via direct injection into cisterna magna (guided by computed tomography) through a 23G spinal needle equipped with a T-capped catheter. The tracer was administered in 1.5 ml of a 6.57 mg/ml solution in sterile saline through the T-cap. The catheter line (total internal volume <50 µl) was then flushed with 300 µl of sterile saline.

Both rats and monkeys were euthanized 24 hours after the IT administration of fluorophore labeled rhNAGLU with 100 mg/kg of pentobarbital IV under isoflurane anesthesia.

In two rats and in both monkeys, the CNS vasculature was counterstained with Texas Red. The animals were heparinized 3 minutes before the euthanasia. The upper body was perfused from the aortal arch to superior vena cava with 5% dextrose in lactated Ringer solution (in one of the monkeys, the perfusion was carried out in the opposite direction to stain predominantly veins). The perfusion was carried out at 7.5±0.5 mm Hg until complete displacement of the blood. Then, Texas Red, 20 mg/ml in dry acetonitrile, was injected into the inflowing perfusate (50 µl per rat, 500 µl per monkey). The perfusion continued to fully wash out the unreacted dye (ca. 20 ml in rats, 500 ml in monkeys).

Cryosectioning. After the euthanasia, the brain and cervical spinal cord were removed and sectioned coronally into three (rats) or five (monkeys) segments: front, middle and cerebellum with cervical cord. The segments were fast-frozen on a 0.5" thick aluminum alloy 6061 block (Grainger, Lake Forest, IL, cat. # 1NYP5) hydrophobised with silicone oil and cooled on dry ice, and embedded in O.C.T compound (Tissue-Plus 4585, (Scigen Scientific Gardena, CA).

The frozen segments were kept at -21°C and cryosectioned on a Leica CM3050 S cryostat (Leica Biosystems, Buffalo Grove, IL) into coronal (in some instances sagittal) 20 μm slices (CT: -15° C; OT: -12° C) and 100 μm coronal slices (CT: -11° C; OT: -10° C). The cryosections were placed on 25.4x76.2x1mm microscopy slides, air-dried, and covered with 0.17 mm thick cover slips over OCT.

**2.4  Photoimaging.**

The microscopy slides were imaged on a Nikon Eclipse epifluorescence TE300 microscope equipped with long working distance objectives (CFI Plan Fluor ELWD 20X and 60X, Plan Fluor 4X and Plan Achromat 1X). Narrow band filter cubes (high throughput filter sets # 49000, 49009, 49011 from Chroma Technology, Bellows Falls, VT, USA) and Nikon DS-Qi2 photon counting camera were used for image acquisition. The dark current of the camera was 395±5 photons•$s^{-1}$ per well. The three fluorescence channels were well isolated from each other, with no detectable FITC crosstalk to the blue and red channels, no detectable Alexa 350 crosstalk to the green and red channels, no detectable Texas Red crosstalk to the green channel and 0.16±0.21% Texas Red crosstalk to the blue channel.

The images of the CNS cryosections were acquired in all three fluorescence channels, in phase contrast (X20 and X60) and in bright field with crossed polarizers (X4). Fluorescence background in all three channels was registered every day using a covered microscopy slide with a layer of OCT. The images were processed using Nikon Elements AR 4.30.02 software.

For each studied CNS cryosection, a series of 4X frames completely covering the slice was acquired and merged in one image. Then, 20x and 60x images were acquired from multiple regions.

Sections obtained from the areas corresponding to plates ## 3, 25, 51, 87, 115 and 161 of the Paxinos atlas[51] were used for counting the perivascular channel entering the parenchyma in rats. In monkeys, the channel counting was performed in sections made through the middle of the brain, approximately corresponding to slice 3 in Calabrese et al[52], Fig. 3. Identification of sulci was in accordance with Von Bonin and Bailey[53].

Images were used for analysis without further processing.

Images intended for deconvolution were acquired as 3D stacks (2 μm per sub-slice) using Nikon Plan Apochromate λ 60X Oil objective and processed using Nikon EDF or 3D deconvolution algorithms.

For presentation, images were exported to monochrome TIFF format. Monochrome TIFF images were pasted into the respective layers of RGB templates to obtain mono- or multicolored images. In all images and individual RGB channels the LUTs are linear and cover the full range of the data.

Surface intensity plots were produced using Nikon Elements AR software.

**2.5  Photoimaging data analysis**

The general patterns of the tracer distribution were studied in the whole-slice composite images manually stitched from 1X (monkey) or 4X images. The detailed patters of perivascular architecture was investigated in 20X and 60X images from 20 μm cryosections.

Cryosections with counterstained vasculature were used to establish the relative positioning of the blood vessels and cells accumulating the tracer. The preliminary estimates of the perivascular channel counts in various CNS sections were obtained from images of 20 μm cryosections and multi-slice image stacks. Only channels with patterned linear probe deposition were counted at their origins at external and internal boundary layers of the CNS. Patterned perivascular cells were clearly identifiable (signal to background ratio = 12000±450:50±20 photons•s$^{-1}$ per well) and distinguishable from the surrounding parenchyma (pattern to the parenchymal cell ratio = 12000±450:250±100 photons•s$^{-1}$ per well).

Considering the out-of-plane positioning of some channels, which prevented their recognition in the thin 20 μm sections, the final counting was carried out in 20X images of 100 μm sections. The numbers of the characteristic linear patterns formed by the perivascular cells per unit of length of the respective CNS boundary were calculated and then multiplied by the section thickness to obtain channel counts per unit of the boundary area.

The same counting technique was applied to the experimental (FITC-rhNAGLU injected) and control groups of rats, n=6 each. Channel counts were obtained from 248 experimental and 241 control cryosections. Counting was carried out by four investigators, three of which participated in studying both experimental and control images. As compared to the control group, this counting approach provided >99.5% specificity calculated as ratio of channel patterns identified in the experimental and control groups, 4247 and 19 respectively. All 19 false positive counts were from two control animals out of six, 15 from one rat and 4 from another.

The channel origination densities (channel•mm$^2$) obtained for each specific CNS boundary were tabulated, and mean counts and standard deviations were calculated for the experimental and control groups.

In cryosections obtained from non-human primates, channel counting was carried out analogously at the external CNS surface and in sulci and fissures.

All original imaging files in Nikon nd2 format (photoimaging), Siemens MicroPET format (PET imaging) and the numerical data are available upon request.

**2.6  Limitations**

The limitation of the method relates to the lack of full statistical information on the existence of macromolecule – transporting perivascular conduits not populated by perivascular macrophages. If such channels do exist, they were not counted. Analysis of images obtained with co-staining of the arteries in rodents (perfusion staining with Texas Red) showed that < 1% of the stained perivascular channels extending from the CNS surfaces may be unpopulated (2 arteries out of counted 300). Therefore, although the densities of the perivascular entrances reported in this paper may be somewhat underestimated (most likely, by not more than single percents), this does not affect the conclusions of this report.

Small perivenous channels traced from large, morphologically identifiable veins were found to contain labeled perivascular cells at the very entrances but no highly labeled perivascular macrophages were identified deeper in the parenchyma. The available data at this point do not allow to determine whether there is no significant transport of macromolecules up the perivenous spaces or there are no cells taking up the tracer in the deep perivenous channels.

## 3  RESULTS

Being a lysosomal protein, rhNAGLU was sufficiently stable after internalization by cells to study its deposition in the CNS cells 24 hours post administration (Supplement Figure S1), i.e., at the time when the free protein has been already cleared from the CSF (Figure S2). Consequently, the background fluorescence was minimal and the tracer deposition patterns were readily discernible in the CNS of rats and monkeys.

### 3.1  Rats.

The highest levels of tracer accumulation were found in cells lining the CNS surface (pia mater and its extensions into the fissures and other internal boundaries) and in some perivascular cells (Figure 1-3). In these cells, fluorescence exceeded the emission of the surrounding cells by up to two orders of magnitude (see Methods), resulting in high-contrast images of cells and structures accumulating the tracer. The intracellular vesicular pattern of fluorophore deposition (Figure S3) was in agreement with endocytosis as the expected mechanism of cell labeling.

Essentially the same levels of tracer deposition were found in the cells lining the internal (intrafissural) and external CNS boundaries, except for the boundaries of the ventricular system. The internal boundaries were homogenously fluorescent throughout their entire depth (Figure 1).

Perivascular spaces were found to extend not only from the external but also from the internal boundaries, excluding the ventricular system (Figure S4). The perivascular spaces around larger (d>0.1 mm) vessels morphologically identifiable as both arteries and veins were surrounded by nearly continuous layers of fluorescent cells (Figure 2), in which the tracer concentration was nearly as high as in the pial cells. No tracer accumulation was observed within or on the luminal side of the walls of vessels of any type or size.

Smaller perivascular spaces had a characteristic linear "dotted" pattern of tracer accumulation at their origins (Figure 3a) first reported by Wagner et al[31] (Figure 3b). Such patterns, entering the parenchyma at angles $70°$-$90°$ to the boundary plane (Figures 1, 3, 4 a), were readily identifiable in all CNS regions except the pineal gland and fissures of the cerebellum.

Co-staining of blood vessels with Texas Red revealed that the regularity of the linear "dotted line" pattern completely disappeared with vessel branching. Some of the perivascular spaces had single or irregularly distanced tracer-accumulating cells even at their origin (Figure 4). In the deeper parenchyma, the regular linear patterns were replaced by irregularly positioned tracer-accumulating cells. Investigation of sections with co-stained blood vessels showed that most of these cells had a perivascular location (as in Figure 4 c, 4 d). Other cells possibly belonged to out-of-plane perivascular channels (Figure 5a).

In the olfactory bulbs (Figure 1a, 5a), 68±24 perivascular channels per $mm^2$ were entering from the ventral surface, 28±18 $mm^{-2}$ from the dorsal surface, and 107±14 $mm^{-2}$ from the boundary between the bulbs. Deeper in the bulb, tracer accumulation was primarily in scattered cells, gradually decreasing in numbers from the glomerular towards the granule cell layer, but present even within the most medial regions.

In the frontal sections of the brain, the highest density of perivascular channels was in the middle cleave of basal forebrain (174±145 $mm^{-2}$) and in the longitudinal fissure (90±19 $mm^{-2}$). The basal forebrain outside of the cleave had 40±17 channel entrances per mm$^2$. The dorsolateral and ventrolateral surfaces of the frontal cortex had 31±14 and 27±12 labeled channel entrances per mm$^2$, respectively. In a section cut through the rostral anterior commissure (Figure 1 b), linear conduits were observed in the cortex, cingulate cortex, and basal forebrain. Labeling of scattered individual cells was also noted within the corpus callosum and the deep parenchyma of internal brain structures. Both scattered and linear-patterned tracer accumulation was especially remarkable within the basal forebrain.

Farther caudally, linear-patterned and scattered tracer accumulation was present in the neocortex, hippocampal formation, thalamus, and hypothalamus, as well as in other internal brain structures (Figure 1c). The highest densities of perivascular entrances, 49±7 mm$^{-2}$ and 55±16 mm$^{-2}$, were found in the longitudinal fissure and the ventral surface of the hypothalamus, respectively. In the subhippocampal cistern extending from the quadrigeminal cistern between thalamus and hippocampus (Figure S6), the observed channel density was 26±243 mm$^{-2}$. The dorsal and ventral surfaces of the cortex had densities of 34±16 and 28±10 mm$^{-2}$, respectively. Scattered fluorescent cells were present within internal brain structures containing predominantly neuronal bodies (striatum, thalamus, hypothalamus, globus pallidus), as well as, to a lesser extent, within axonal structures (corpus callosum, corticofugal pathways, optic tracts, stria medullaris of the thalamus, and fimbria of the hippocampus). Remarkable linear-patterned tracer deposition was noted at the boundaries between the dentate gyrus and the superior aspect of thalamus and between the fimbria of the hippocampus and the superior aspect of thalamus. Linear patterns were also observed in areas surrounding larger blood vessels and on the inferior aspects of optic tracts.

In the areas of the caudal cerebrum and rostral cerebellum (Figure 1d, 1e), both linear-patterned and scattered tracer deposition in the neocortex, entorhinal areas, subiculum, brainstem nuclei and white matter were evident. Continuous cell layers accumulating the tracer were observed at the boundaries between the brainstem and pineal gland, the brainstem and cortex, and between the cerebellum and brainstem.

Scattered cells containing the tracer were present in large numbers throughout the pineal gland and the interpeduncular nucleus (Figure 5 b). However, no traceable linear patterns were observed, which may be related to the architecture of the vascular network of these structures that differ from the rest of the CNS[54].

In the midbrain cistern, which in rats is essentially a tight subcortical fissure), channel density was higher than on the external dorsal and ventral surfaces of the caudal cortex than on the surface of the thalamus, 43±17 and 41±17 mm$^{-2}$ vs. 27±13 and 28±11 mm$^{-2}$, respectively.

In the cerebellum (Figure 1e, 4d, S5), continuous layers of labeled cells within its fissures were evident. Most of the vessels extending from the fissures through the granular layers towards the Purkinje layers had only 1 or 2 perivascular cells. Some of the branches that they formed in the granular layer had tracer-labeled perivascular cells as well (Figure 4 d). In the cerebellum, the densities of perivascular entryways on the dorsal and ventral surfaces were 22±12 and 17±11 mm$^{-2}$, respectively. In the brainstem, channel densities on the ventral and dorsal (facing cerebellum) boundaries were 35±12 and 39±16, respectively.

In the spinal cord (Figure 1 f), remarkable linear-patterned labeling along multiple radial channels was present within the lateral white matter. The average channel densities entering from the dorsal and ventral surfaces were 20±13 and 29±13 mm$^{-2}$ respectively. The dorsal and ventral median sulci were covered by highly labeled cells and contained 63±66 mm$^{-2}$ and 59±50 mm$^{-2}$ channel entries, respectively. Labeled cells were scattered throughout the gray matter of the spinal cord as well, but without detectable linear patterns.

Overall throughout the CNS, the highest density of perivascular entryways, 90±71 mm$^{-2}$, was found in the longitudinal fissures and cleaves, whereas at other internal boundaries the channel density was 39±18 mm$^{-2}$. At the external boundaries (not including olfactory bulbs), the perivascular entryway density was 28±15 mm$^{-2}$ (Figure S7). We note that the densities reported in this paper include only linearly patterned channels identifiable with >99% specificity as compared to the control group (Methods).

Although tracer penetration from the perivascular channels to the parenchyma was not the focus of this study, indications of it were clearly present in all studied CNS tissues of rats. Fluorescence patterns and intensities outside of the perivascular spaces indicated a statistically significant exit of the tracer from the perivascular channels into the parenchyma (Figure S8). In a group of animals where fluorescent rhNAGLU was administered systemically (intravenously), no patterns indicating protein accumulation in any particular cells or cell organelles were detected in the CNS (not shown). Therefore, the fluorescence patterns resulting from tracer administration to the CSF cannot be attributed to rhNAGLU uptake from the systemic circulation after tracer clearance from the CSF to the blood.

### 3.2 Non-human primates.

Investigation of cryosections from *M. mulatta* (N=2) has shown that the tracer follows generally the same route as in rats, from the CSF into the internal boundary layers (all sulci and fissures) and perivascular channels. The latter were readily identifiable by the same characteristic linear patterns of tracer endocytosing cells as in rats, with channels originating inside fissures and sulci as well as on the external surface (Figure 6).

Dynamic PET data[55] demonstrated that rhNAGLU (labeled with $^{89}$Zr) concentration in the fissures, sulci and major periarterial conduits (Virchow-Robin spaces) equilibrates with that of the cisternal CSF within 30 minutes after the tracer administration to the CSF. Tracer penetration into the parenchymal compartments (outside of fissures, sulci and major perivascular spaces) continues at a lower rate, reaching maximum at 3±2 hours. The fast equilibration of the tracer concentration between the CSF and the intrafissural/sulcal liquid layers is likely a result of the same biomechanical process (pulsation-assisted remixing/Taylor diffusion) that equilibrates solute concentrations throughout the CSF.

Thus, the in primates, in spite of the much larger dimensions of the brain and other parts of CNS, the perivascular liquid system can perform essentially the same transport functions as in rats. Based on the limited dataset (36 transversal full-brain/spinal cord sections from two animals), we estimate that channel entrance densities at the external and internal CSF surfaces vary within the range of approximately 20 to 100 mm$^{-2}$, with the average for the dataset of 36±22 mm$^{-2}$. No channels were found to be extending from the inner surfaces of the ventricles (although scattered tracer accumulating cells were observed in the vicinity of ventricles as far as 0.2-.03 mm from the surface). A larger study is needed to quantify the channel origination densities in a statistically significant manner.

# 4 DISCUSSION

The ubiquity of the perivascular channels accessible for macromolecules administered to the CSF (in particular, their entrance to the parenchyma from the internal as well as external CNS boundaries) suggests that the perivascular route can potentially provide a robust avenue for delivery of biopharmaceuticals and other macromolecules to the CNS. It is especially encouraging that in primates perivascular transport of macromolecules to the CNS occurs not only from the external surface, but also from the internal boundaries, from which perivascular spaces extend into deeply located compartments.

The mechanism of the fast solute entrance into the (thin) internal liquid layers communicating with the CSF has not been studied. Massive tracer entrance deep into the fissures cannot be explained by diffusion; it seems logical to suggest that the mechanism of solute transport is active. Both arteries and veins run within these layers, hence liquid uptake into the fissures could not be driven by a pressure gradient between the periarterial and perivascular spaces. Pulsatile remixing of the liquid content of the fissures, not unlike that of the CSF in the leptomeningeal space, seems to be the only viable explanation of the observed solute transport into the fissures from the CSF.

The data further suggest that macromolecules enter the perivascular spaces surrounding not only arteries, but also morphologically identifiable (larger) veins. In view of the significant similarities in the arteriole and venule morphology at high branching levels, protein entrance into small perivenous spaces could not be positively confirmed within this study and requires further investigation.

The data strongly support the accessibility of CNS to macromolecular and supramolecular therapeutics delivered to the entrances of perivascular channels leading to the target region. Previously we have shown that such delivery is possible throughout the entire cerebro-cervical and spinal CSF pools[33,34,35,43].

Detectable labeling of non-perivascular cells confirms that perivascular conduits do communicate with the parenchymal interstitial fluid, enabling the transport of macromolecules of at least the size of rhNAGLU (ca. 12 nm for dimer[56]) to the interstitial space of the CNS. The latter represents about 20% of the CNS volume[57] and can provide avenues for further transport of macromolecules to distant cells.

However, in spite of the fenestrated nature of the terminal perivascular sheaths[58], the route as a whole is not entirely barrier-free: the actively endocytosing perivascular cells that we used in this study to trace perivascular channels and cells lining the external and internal boundaries of the CNS are positioned along the route and may intercept any solutes that can bind their endocytosis-mediating receptors. A study on the receptor specificity of these cells would enable developing therapeutics that can reach the parenchymal cells more efficiently.

The similarities in the configuration and densities of the perivascular conduits transporting macromolecules in primates and rats suggest that rodents are potentially suitable for modeling the transport of intrathecally administered macromolecular and supramolecular therapeutics. However, quantitative evaluation of all stages of the transport process, especially the stages of entrance from the perivascular space to the interstitial space (of which little is known) is necessary for direct scaling.

In conclusion, our data support the hypothesis that perivascular conduits can transport macromolecules to most, if not all, subcompartments of the CNS. Further mechanistic investigation of the physiological transport processes in the CSF and in perivascular space, as well as investigation of the functions of the perivascular cells, are required to develop therapeutics optimized for IT administration. Investigation of the above factors will also benefit our understanding of the transport and clearance of endogenous products of CNS[59,60] and interstitially (CNS) administered therapeutics[61].

## 5 Conflict of Interest

The authors declare that the research was conducted in the absence of any commercial or financial relationships that could be construed as a potential conflict of interest.

## 6 Author Contributions

B.D. Performed most of the experimental rat studies, including dissections, cryosectioning and photoimaging; optimized experimental techniques and methods of data analysis; designed data analysis strategies; co-wrote the manuscript (with MP).

J.A. Performed rat studies, cryotomy, photoimaging and data processing in pilot and preliminary studies and for the main experimental and control groups of rats; participated in manuscript preparation.

N.G., P.G. and F.L. Performed cryotomy, photoimaging and data processing for the control group of rats and animals with perfusion-stained vasculature; assisted with photoimaging data processing. Performed experimental studies in non-human primates, including cryosectioning and photoimaging.

V.B Assisted with the intrathecal techniques and data processing; lead PET studies, participated in manuscript preparation.

E.M. and M.H. Provided critical methodological assistance in the experimental studies with non-human primates.

M.P. Conceived and directed the study, designed experiments, performed protein labeling and some of pilot rat studies, carried out perfusion staining of blood vessels, analyzed and interpreted the data, and wrote the manuscript (with BD).

## 7 Funding

This work was supported by NIH grants R21NS090049 and R01NS092838. The protein, rhNAGLU expressed in chicken eggs, was provided by Synageva Biopharma (presently Alexion). The use of the protein in this study was not related to its potential therapeutic applications.

## 8 Acknowledgments

Authors are grateful to Prof. Roxana O. Carare and Dr. Kimberley S. Gannon for valuable discussions and comments on the manuscript.

We thank our summer students Kassandra Boada, Stepan Levin and Connor Whelan for assistance with cryosectioning, photoimaging and data processing.

We thank the Chiari & Syringomyelia Foundation (Staten Island, NY) for organizing an excellent venue of forums facilitating a highly interdisciplinary discussion of the mechanistic challenges faced by research in the area of cerebrospinal drug transport.

The tracer protein, rhNAGLU expressed in chicken eggs, was provided by Synageva Biopharma, Lexington, MA (presently merged with Alexion, Boston, MA). The use of the protein in this study is unrelated to its possible prospective clinical applications.

## 9 Supplementary Material

The full version of this Article has Supplementary Materials.

## 10 Data Availability Statement

The raw data supporting the conclusions of this manuscript will be made available by the authors, without undue reservation, to any qualified researcher.

**FIGURES**

**Figure 1.**

Deposition of FITC-rhNAGLU in the CNS of rat, 24 hours after administration to cisternal CSF. Unstained, unfixed coronal 20 μm cryosections. a. Olfactory bulbs. b. Brain section at the level of rostral anterior commissure. c. Brain section through hippocampal formation and thalamus. d. Brain section through pineal gland. e. Cerebellum and brainstem section ca. 0.5 mm rostral of the facial nerves. f. Section through cervical spinal cord. See Methods for section reference to Paxinos atlas[51].

Field of view: A: 8.8x5.9 mm, B-E: 17.7x11.8 mm, F: 4.4x2.9 mm

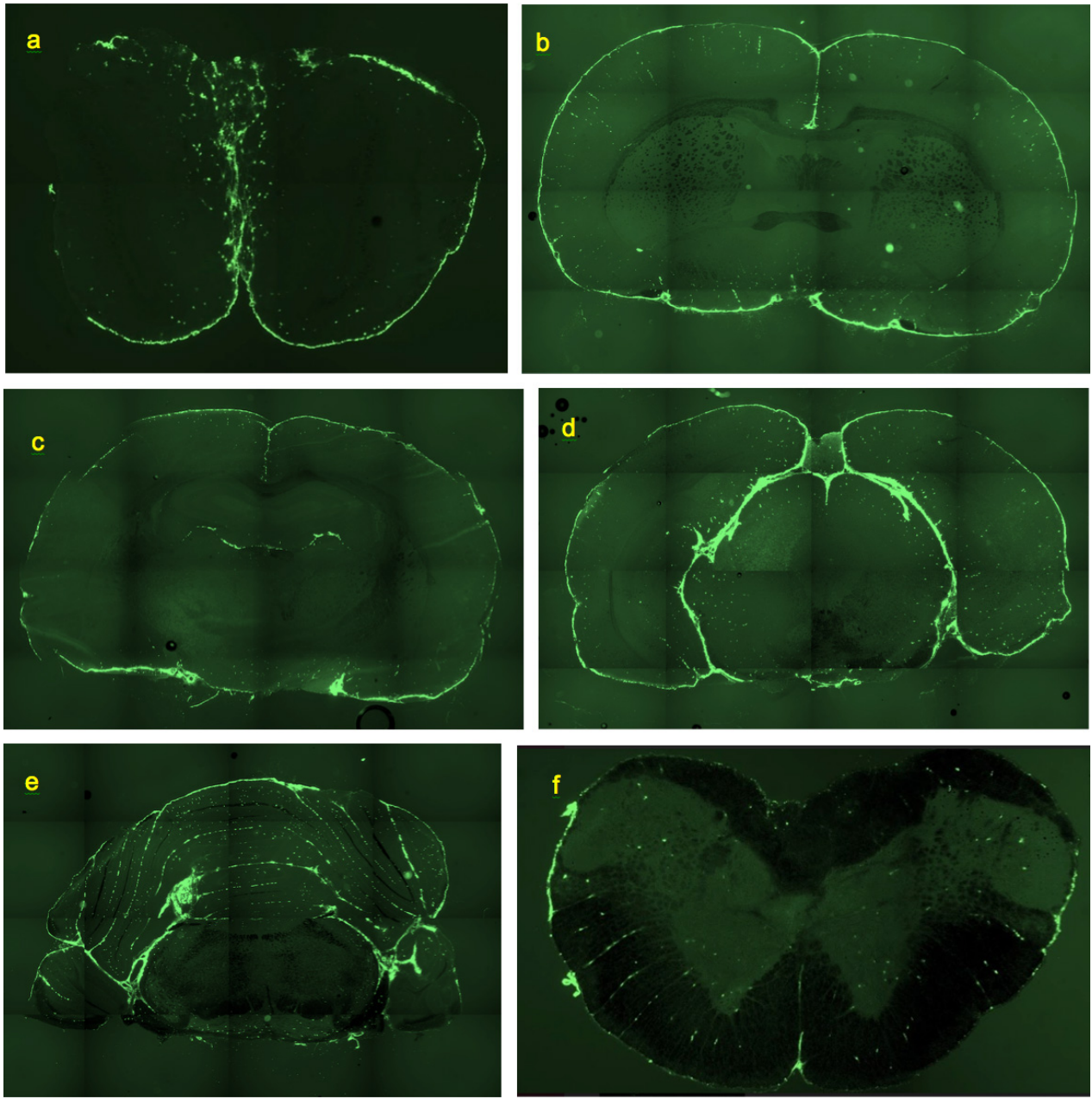

**Figure 2.**

FITC-rhNAGLU accumulation (green) around major blood vessels (red) in the perivascular spaces extending from the subhippocampal cistern. Unstained unfixed 20μm sagittal cryosection 3 mm off the central plane, 0.9 x 0.6 mm (See Figure S 6 for the anatomical context).

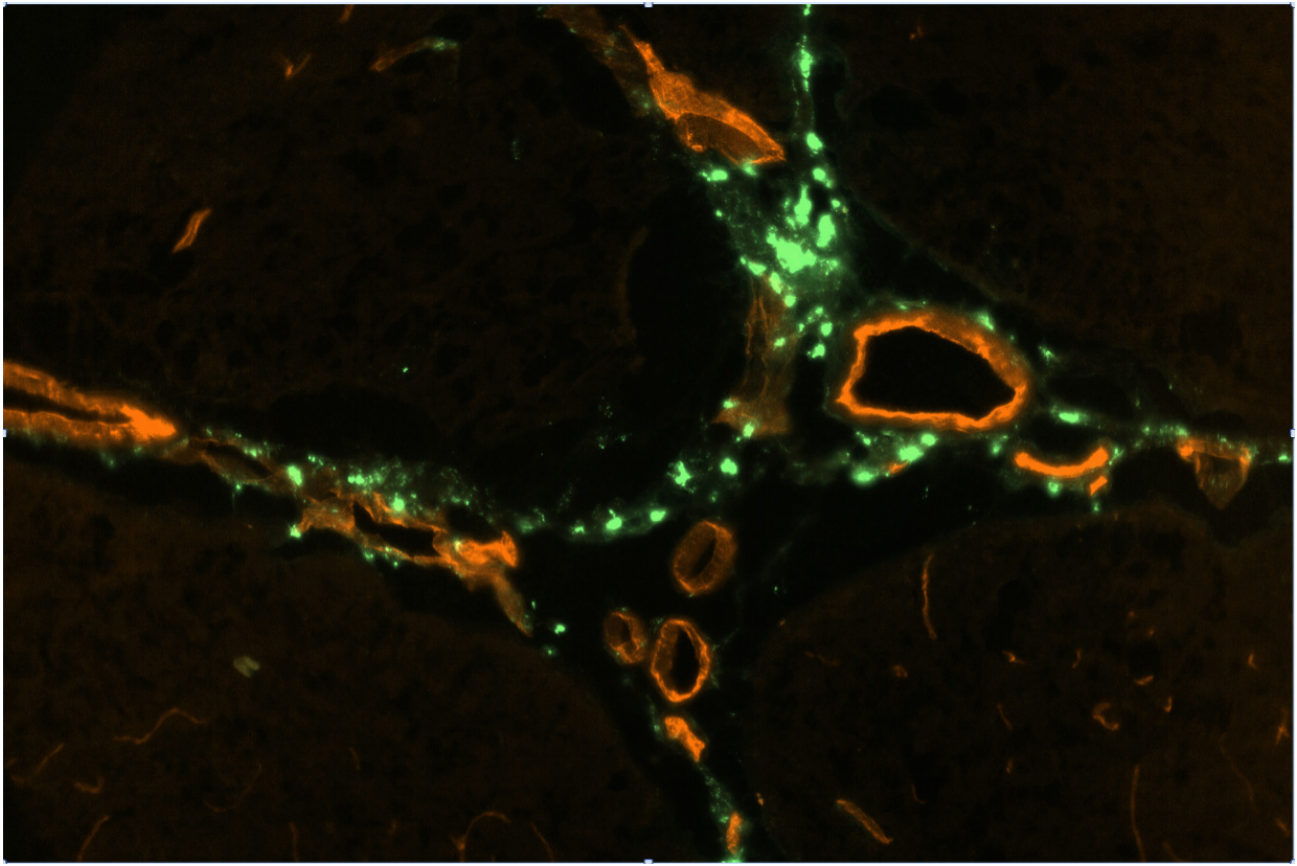

**Figure 3.**

a. Small regularly labeled perivascular space extending from the surface of the frontal cortex, 0.6x0.4 mm.  b. From [31]: analogous perivascular space, peroxidase/3,3'-diaminobenzidine staining. c. Irregularly labeled perivascular channel in the cerebellum, 0.9 x 0.6 mm. a, c: unstained unfixed 20μm coronal cryosections. Red: blood vessel extending from a cerebellar fissure.

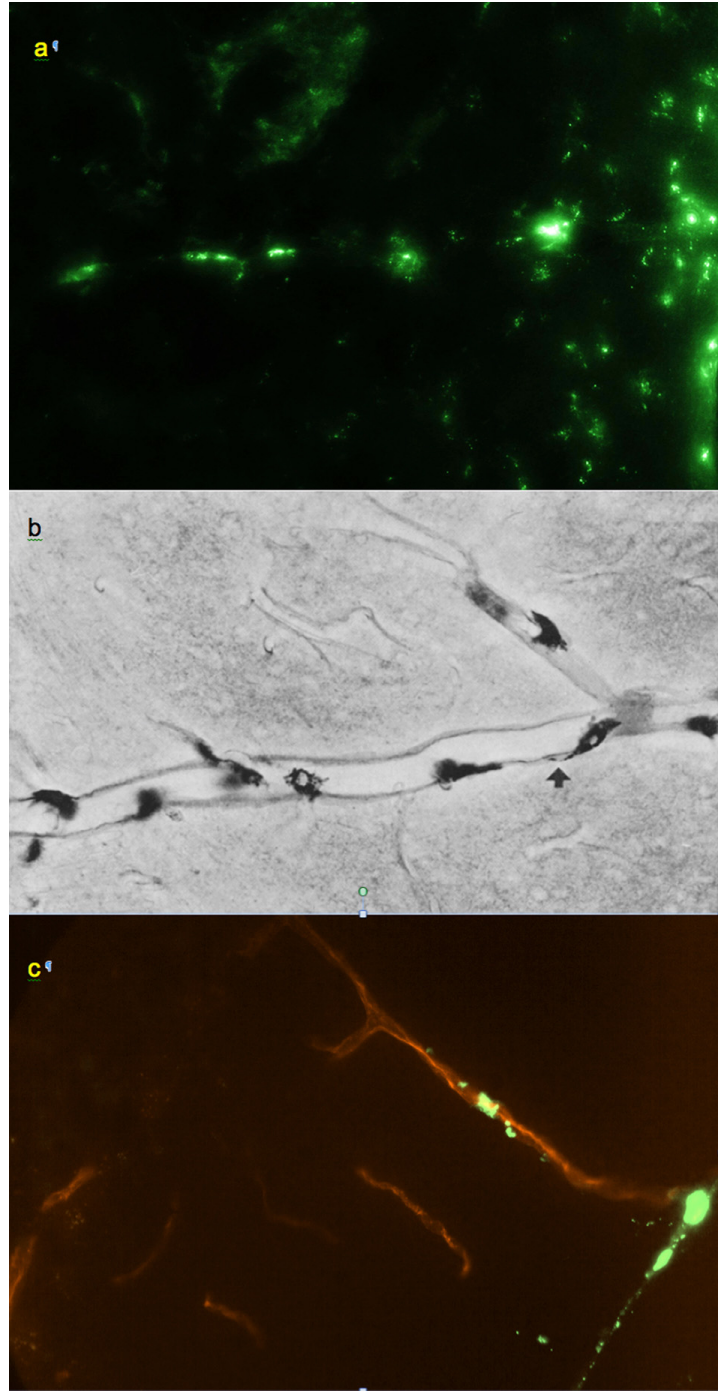

**Figure 4.**

Morphology of perivascular conduits (green: FITC-rhNAGLU, red: blood vessels stained with Texas Red, cyan: cell nuclei stained with Hoechst 33342). a. Frontal cortex, vessel entrance from the external surface of the brain; 0.6 x 0.4 mm, sagittal plane.  b. Conduits (horizontal linear patterns) extending from the longitudinal fissure (vertical line in the center), 0.9 x 0.6 mm, coronal plane. c. Scattered labeling of internal perivascular spaces in caudate putamen; sagittal plane, 0.3x0.2 mm. d. Cerebellar fissure and perivascular spaces extending through the molecular cell layer to the Purkinje cells and granular cell layers, 0.9 x 0.6 mm, coronal plane. Unstained, unfixed 20 μm cryosections.

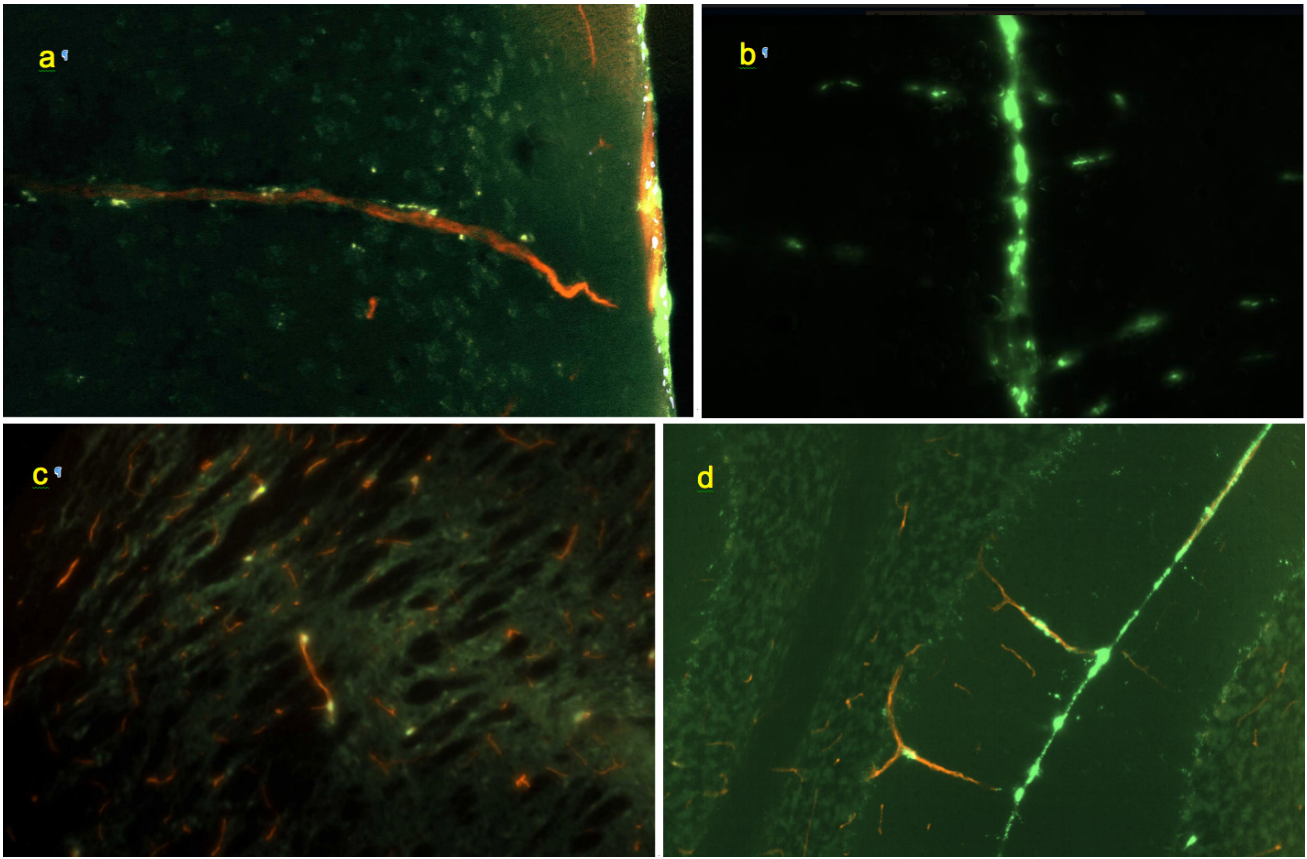

**Figure 5.**

a. Perivascular channel at the dorsal surface of olfactory bulb (green); blood vessels stained with Texas Red; 0.9 x 0.6 mm; sagittal slice.  b. Pineal gland: non-linear pattern of tracer deposition; 0.9 x 0.6 mm, transverse slice; no co-staining.

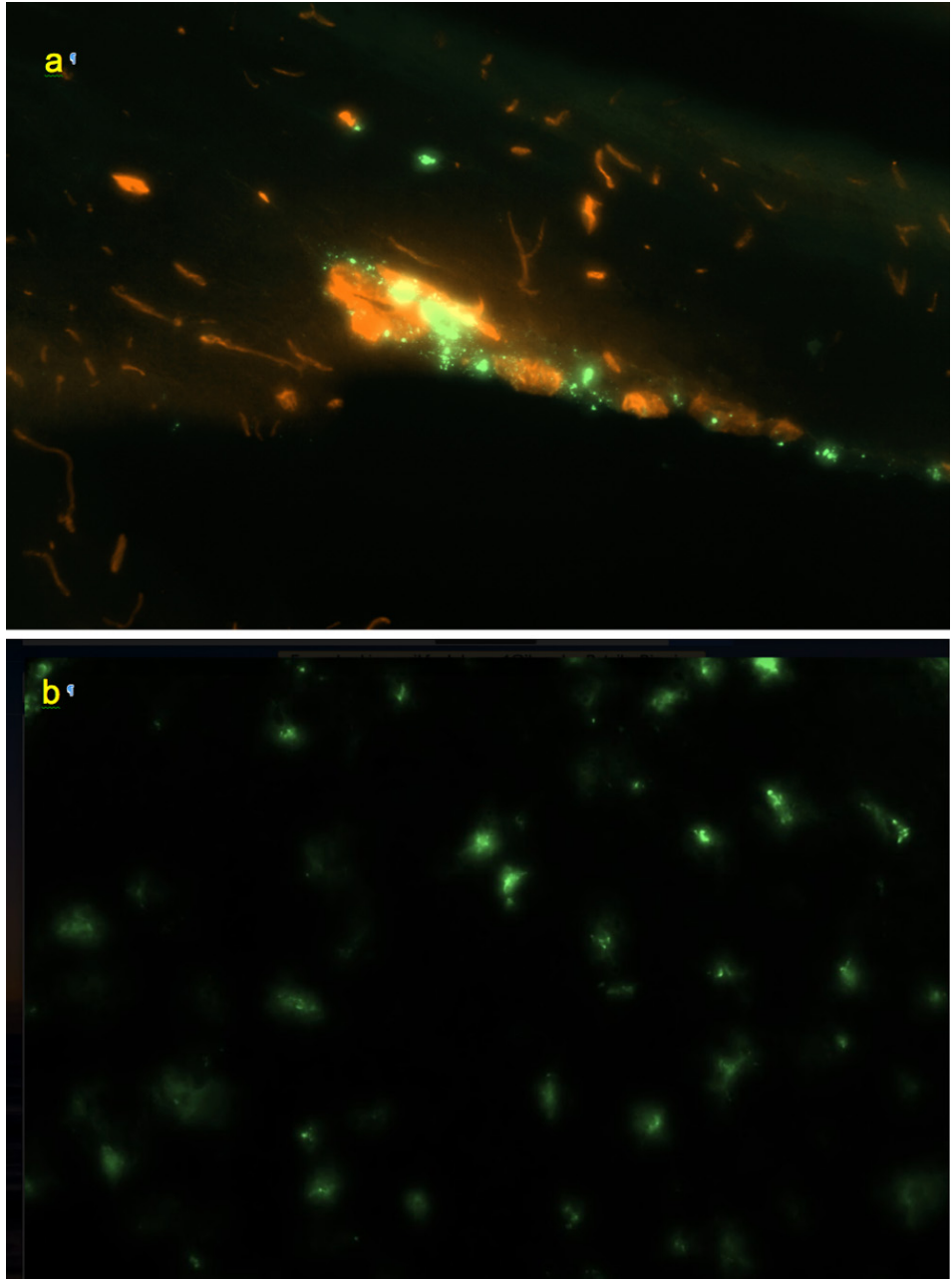

**Figure 6.**

Labeling of perivascular channels with Alexa-350-rhNAGLU in monkey (M. fascicularis). a. Perivascular cells around arteries branching from the arcuate sulcus (AS) and entering cortex from its surface (CS), 4.4x2.9 mm.  b. Perivascular channels branching from the anterior subcentral sulcus (SAS), 3 x 2 mm. 100 μm cryosections.

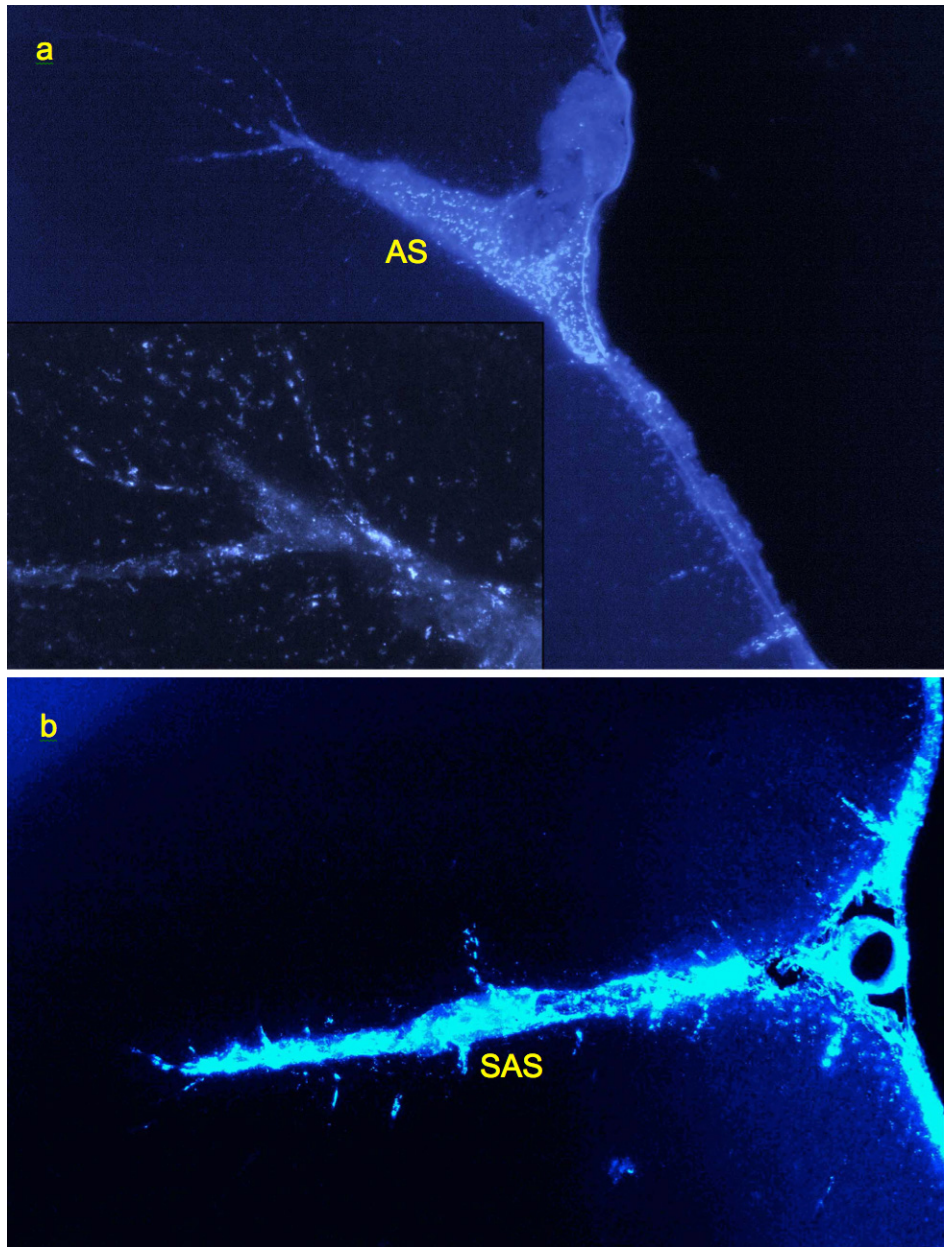